\documentclass{iau} 
\usepackage{graphicx}
\usepackage[authoryear]{natbib}

\newcommand{\msun}{M$_\odot$}
\newcommand{\rsun}{R$_\odot$}
\newcommand{\kms}{km~s$^{-1}$}

\newcommand{\mytilde}{\raise.17ex\hbox{$\scriptstyle\mathtt{\sim}$}}
\def\lesssim{\mathrel{\hbox{\rlap{\hbox{\lower3pt\hbox{$\sim$}}}\hbox{\raise2pt\hbox{$<$}}}}}
\def\gtrsim{\mathrel{\hbox{\rlap{\hbox{\lower3pt\hbox{$\sim$}}}\hbox{\raise2pt\hbox{$>$}}}}}
\def\lesseq{\mathrel{\hbox{\rlap{\hbox{\lower3pt\hbox{$-$}}}\hbox{\raise2pt\hbox{$<$}}}}}
\def\gtreq{\mathrel{\hbox{\rlap{\hbox{\lower3pt\hbox{$-$}}}\hbox{\raise2pt\hbox{$>$}}}}}

\title[Post-common envelope PN] 
{Post-common envelope PN, \\fundamental or irrelevant?}

\author[O. De Marco et al.]   
{Orsola De Marco$^{1,2}$,
T. Reichardt$^{1,2}$,
R. Iaconi$^{1,2}$,
 T. Hillwig$^3$,
 G. H. Jacoby$^4$,
 D. Keller$^5$,
 R. G. Izzard$^6$,
 J. Nordhaus$^7$
 \and E. G. Blackman$^8$}

\affiliation{$^1$Department of Physics \& Astronomy, Macquarie University, Sydney, NSW 2109, Australia \\ email: {\tt orsola.demarco@mq.edu.au} \\[\affilskip]
$^2$Astronomy, Astrophysics and Astrophotonics Research Centre, Macquarie University \\[\affilskip]
$^3$Department of Physics \& Astronomy, Valparaiso University, Indiana, USA\\[\affilskip]
$^4$Lowell Observatory, Flagstaff, AZ, USA\\[\affilskip]
$^5$KU Leuven, Belgium\\[\affilskip]
$^6$Institute of Astronomy, Cambridge University, UK\\[\affilskip]
$^7$National Technical Institute for the Deaf, Rochester Institute of Technology, NY, USA\\
[\affilskip]
$^8$University of Rochester, NY, USA
}

\pubyear{2016}
\volume{xxx}  
\jname{Title of your IAU Symposium}
\editors{A.C. Editor, B.D. Editor \& C.E. Editor, eds.}
\begin{document}

\maketitle

\begin{abstract}
One in 5 PN are ejected from common envelope binary interactions but {\it Kepler} results are already showing this proportion to be larger. Their properties, such as abundances can be starkly different from those of the general population, so they should be considered separately when using PN as chemical or population probes. Unfortunately post-common envelope PN cannot be discerned using only their morphologies, but this will change once we couple our new common envelope simulations with PN formation models.
\end{abstract}

\firstsection 

\section{The fraction and chemistry of post-common envelope PN}
At least one in 5 planetary nebulae (PN) derives from a common envelope (CE) binary interaction, where the PN is the ejected CE and the inner close binary is the result of the companion's in-spiral into the progenitor AGB stars \citep{Ivanova2013}. This fraction is a lower limit because we cannot detect photometric variabilities $\lesssim$0.1~mag, due primarily to poor sampling of ground-based observations. 
A search carried out with the {\it Kepler Space Telescope} resulted in the detection of 4  variables out of 5 objects with data \citep{DeMarco2015}. Of the 4 variables, one is a double-degenerate binary central star, and one is likely to be a pole-on binary. With variability amplitudes of 0.0007 and 0.0005~mag, neither would have been detected from the ground. 

A binary search  using data from {\it Kepler II} (K2), campaigns 0, 2, 7 and 11 is underway. Campaigns 0, 2 and 7 include a grand total of 15 PN, two of which have periodic variability consistent with binarity. The variability amplitudes, 0.05 and 0.02~mag, respectively, are at the low end of what could be detected from the ground. However, the sensitivity of K2 is up to 5 times worse than for {\it Kepler} and neither of the 2 binaries detected with {\it Kepler} would have been detected by K2. In addition, 11 of the 15 K2 targets analysed are compact PN, while all the {\it Kepler} targets are extended:  the variability detection threshold is therefore much worse for the bulk of the K2 targets. K2, Campaign 11 has 139 PN on silicon of which 22 are extended. Hopefully those data will return better statistics.
\begin{figure}[b]
\begin{center}
 \includegraphics[angle=-90,width=3.7in]{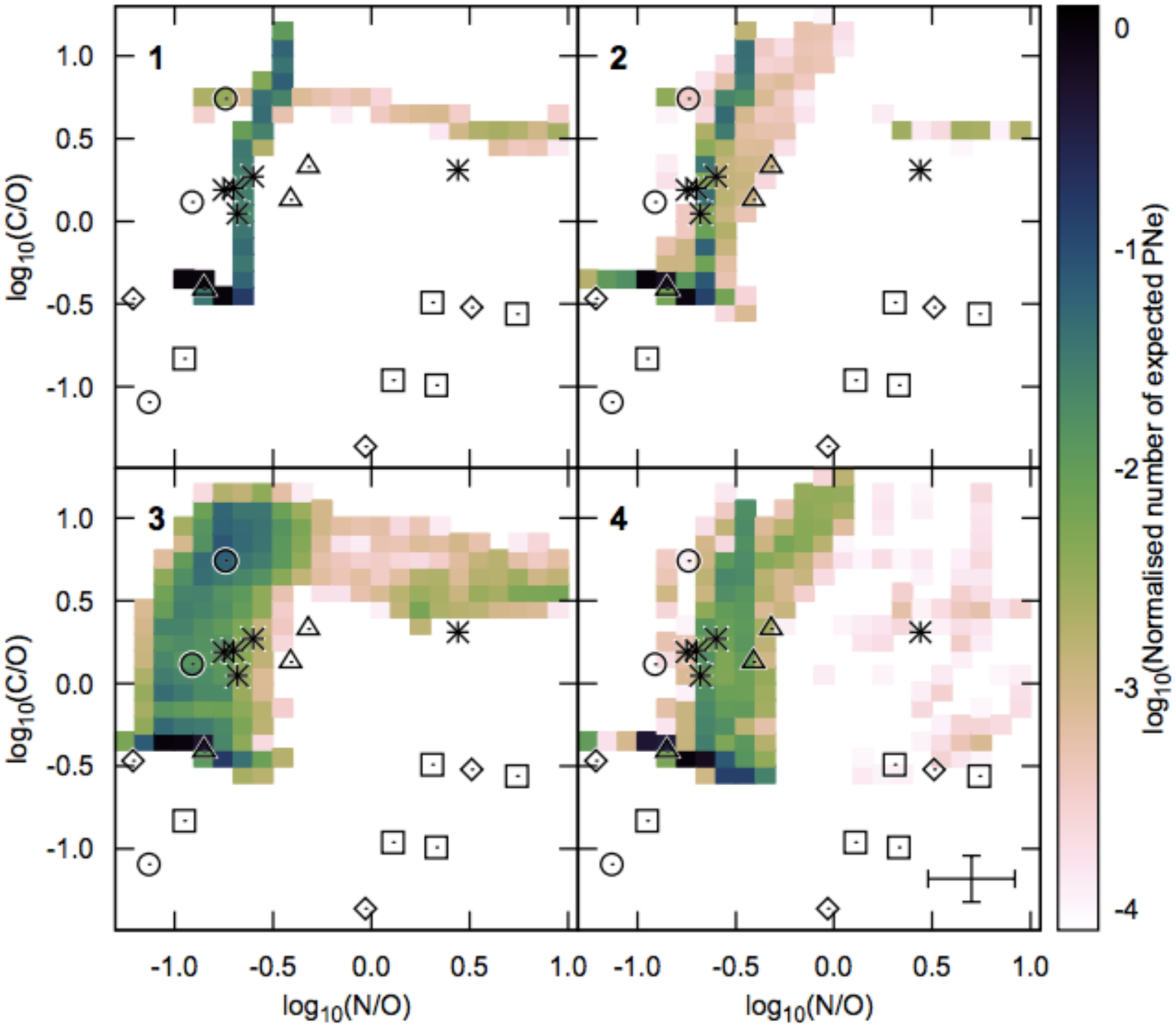} 
 \caption{Observed vs. predicted C/O and N/O ratios for PN produced via different channels: (1) single stars, (2) wind mass-loss from binary star origin, (3) CE merger + wind mass loss, (4) CE ejection. Observation symbols: round: round PN; star: elliptical PN; square: bipolar PN; triangular: elliptical PN with bipolar core, rhomboid: quadrupolar PN. Observations: \citet{Leisy2006,Stanghellini2005}. Figure from Keller, Izzard and Stanghellini, in prep.}
   \label{fig1}
\end{center}
\end{figure}
The {\it Kepler} data show that low amplitude post-CE PN central binaries exist, as predicted and that the fraction of post-CE PN could be substantially larger than the lower limit of one in 5. This lower limit is already larger than the prediction from population synthesis of 9\% for Solar metallicity of \citet[][which are consistent with previous work, e.g., \citet{Yungelson1993,Nie2012}]{Izzard2015}. An over-abundance of post-CE PN could be explained either if some single stars make under-luminous PN \citep{Soker2005} or if some post-CE PN are mimics, such as for example Stroemgren spheres around hot stars \citep{Frew2010}.  

Expectations of the properties of post-CE PN are different from those of PN from other evolutionary channels.  Post-CE PN have a curbed AGB evolution, resulting in reduced C/O ratios and lower s-process abundances. A secondary effect is that more massive stars are in binaries more often, something that could inflate the relative number of more massive stars in the post-CE PN population resulting in relatively larger N/O ratios. In Fig.~2 we can compare the top-left and bottom-right panels where we see the relatively broader distributions in the post-CE channel and also the larger, though still small, proportion of PN at high N/O and very low C/O not present in the single star channel. The observations are still insufficient for a proper comparison, but it is already clear that the observed bipolar PN at low C/O and mostly but not exclusively high N/O (more massive central stars) can only be (approximately) reproduced by the CE channel.

\vspace{-0.5cm}
\section{Observed and simulated post-common envelope PN}
Unfortunately we cannot unambiguously recognise post-CE PN based on their morphology alone. Approximately 50 post-CE central stars are known to date and about two thirds of them are reasonably well imaged. As shown by \citet{Miszalski2009b}, many of these PN are approximately bipolar (e.g., A~41, A~79, K~1-2, Fig.2), but there is some diversity.  Some PN display only a narrow waist (e.g., HaTr~4), sometimes with jets (A~63; jets not visible in Fig.~2). Ring nebulae (e.g., NGC~6337, Sp~1) are likely to be pole-on views of waists. More complex morphologies are also present: elliptical PN with jets (NGC~6778, Fleming~1), more or less elliptical distributions of faint gas on which we see superimposed stark and very collimated double (K~1-2, PN~G~135.9) or single jet-like structures (HFG~1, M~2-29).

\begin{figure}[b]
\begin{center}
 \includegraphics[angle=-90,width=4.5in]{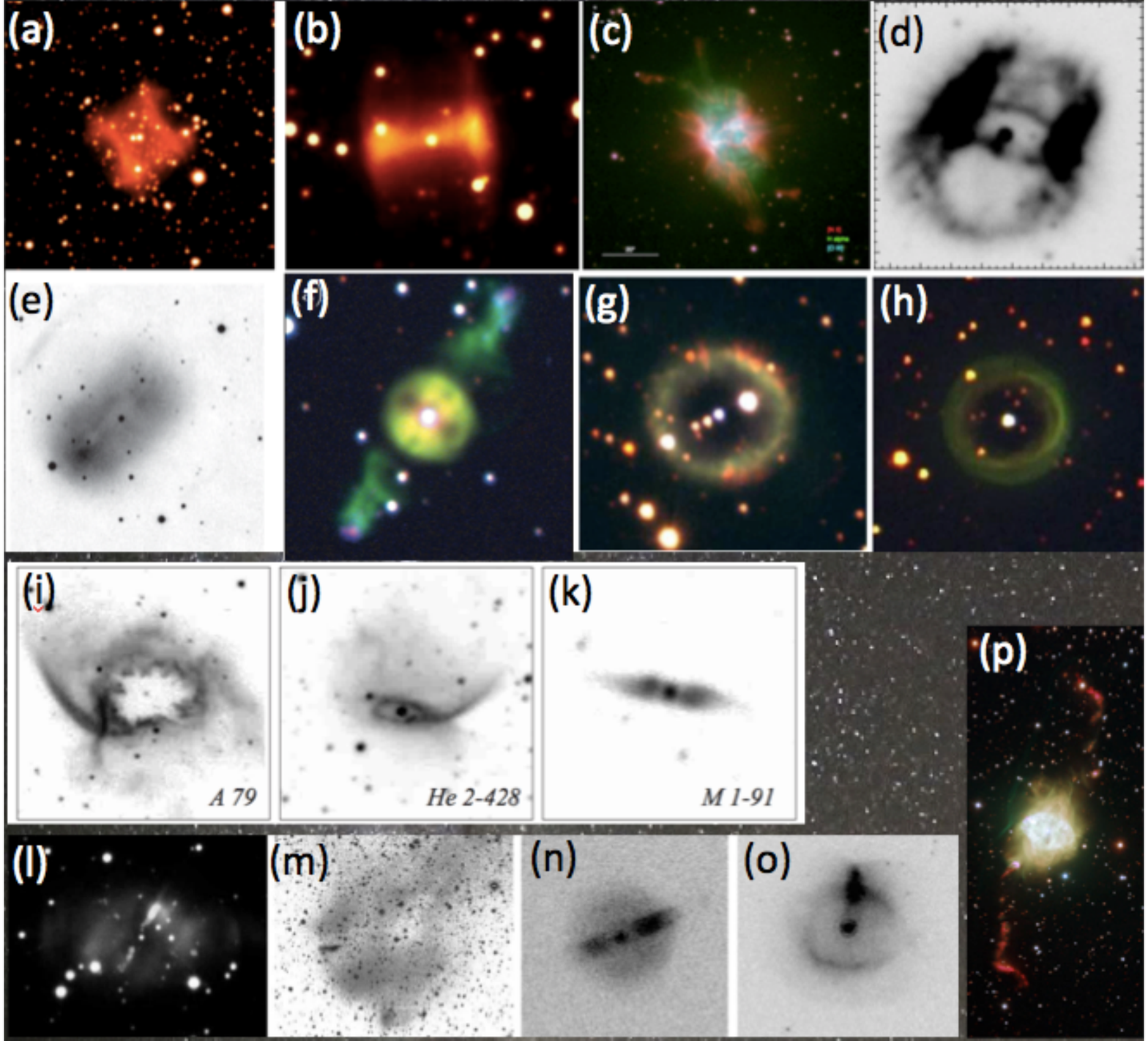} 
 \caption{PN around post-CE central binaries;
 (a) A~63 \citep{Mitchell2007}; 
 (b) HaTr~4 (from the ESO NTT archive); 
 (c) NGC~6778 \citep{Guerrero2012}; 
 (d) A~41 \citep{Jones2010}; 
 (e) A~65 \citep{Huckvale2013}; 
 (f) ETHOS~1 \citep{Miszalski2011}; 
 (g) NGC~6337 \citep{Hillwig2010}; 
 (h) Sp~1 \citep{Hillwig2016b}; 
 (i) A~79, 
 (j) He~2-428, 
 (k) M~1-91 \citep{Rodriguez2001}; 
 (l) K~1-2 \citep{Corradi1999}; 
 (m) HFG~1 \citep{DeMarco2009b}; 
 (n) PN~G135.9+55.9 \citep{Napiwotzki2005}; 
 (o) M~2-29 \citep{Hajduk2008}; 
 (p) Fleming~1 \citep{Boffin2012}.}
   \label{fig1}
\end{center}
\end{figure}
\begin{figure}[b]
\vspace*{-1.5 cm}
\begin{center}
 \includegraphics[angle=-90,width=5in]{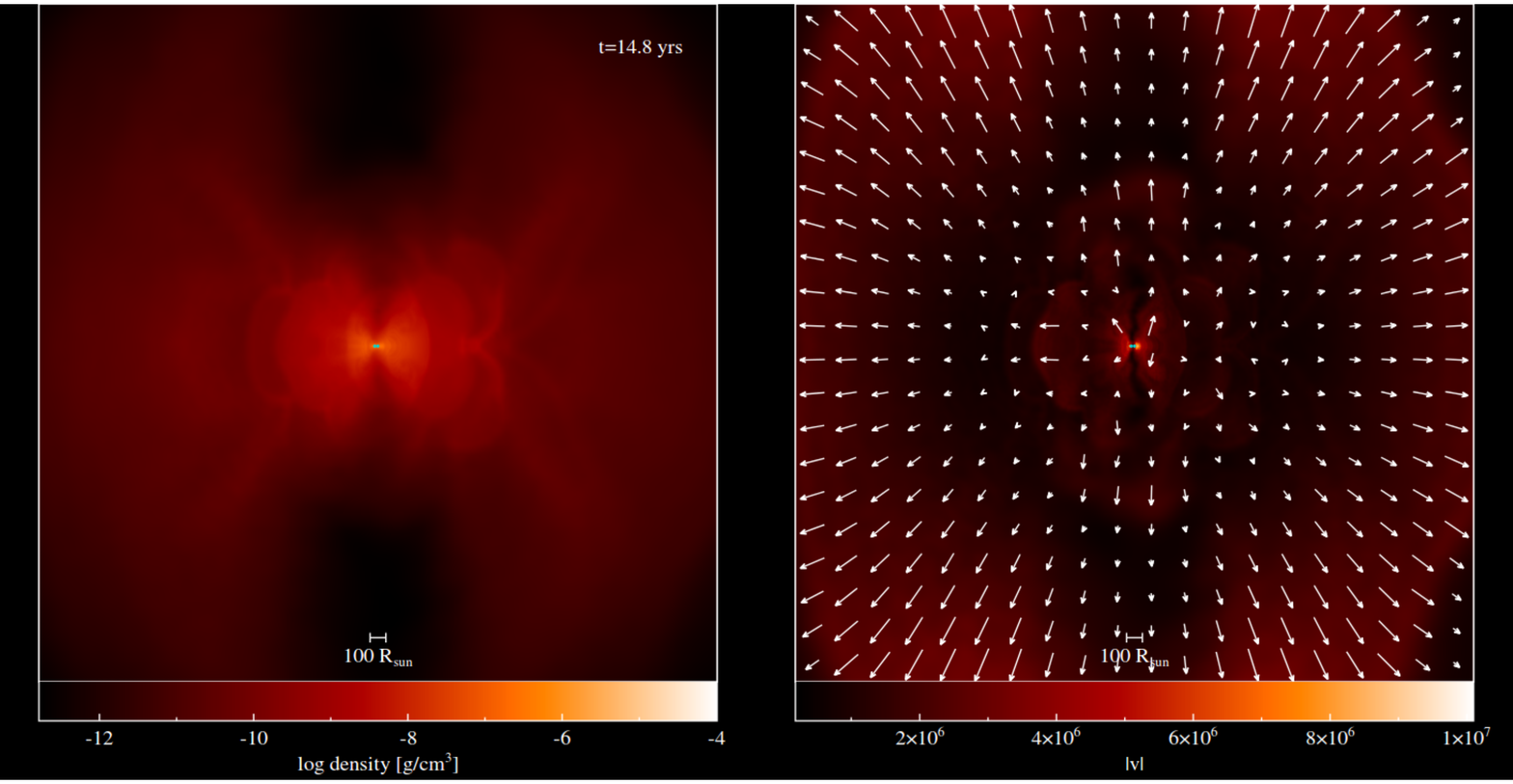} 
 \vspace*{-1.3 cm}
 \caption{Density (left) and velocity (right, in cm~s$^{-1}$; arrows indicate flow direction) slices perpendicular to the orbital plane at 14.8 years of an SPH CE simulation between a 0.88~\msun, RGB star and a 0.6~\msun\ companion, carried out with the {\sc phantom} code \citep{Iaconi2017}.}
   \label{fig1}
\end{center}
\end{figure}

The only way to understand how these shapes result from CE ejections is via simulations, such as those by \citet{GarciaSegura1999} or \citet{HuarteEspinosa2012b}. Nobody to date has simulated a post-CE PN {\it accounting for the common envelope ejection}. Simulations of the CE interaction have shown that the ejection of the envelope takes place primarily on the orbital plane \citep{Sandquist1998,Passy2012}. We have recently carried out a 2-million particle SPH simulation with the code {\sc phantom} \citep[][]{Iaconi2017} and the parameters of Passy et al. (2012), but starting the simulation at the time of Roche lobe overflow, something that allows the distribution of ejected gas to be more realistic. In addition, with SPH the entire gas distribution at the end of the simulation ($\sim$15 years after Roche lobe overflow) is known, albeit with varying resolution (from 0.3~\rsun\ near the binary to 80~\rsun\ near the outer parts of the distribution). 

This simulation has not considered the fast wind, nor the ionising radiation. We could plausibly assume that the central star at the end of our simulation is hotter than $\sim$20,000~K. If so,  both ionising radiation and a fast tenuous wind would start blowing into the dense CE structure. While this has not been done yet in these simulations, using this image we can already make the following considerations. If a fast wind were to start at the end of the CE in-spiral it would meet with incredibly high densities, of the order of ~10$^{-8}$~g~cm$^{-3}$ in the vicinity of the stars (Fig.~3, left panel). Even the evacuated, funnel-like regions seen above and below the equatorial plane have high densities ($\sim$10$^{-11}$~g~cm$^{-3}$). By comparison typical PN simulations blow a fast wind into a much more tenuous medium with typical densities of 10$^{-21}$~g~cm$^{-3}$, e.g., \citet{GarciaSegura2006}. This shows that the CE structure presents more or less a solid wall to the fast wind. On the other hand the entire CE is expanding at a range of speeds with the bulk of the volume expanding at 20~\kms. Hence, in 100-1000 years the densities will decline by 5 to 8 orders of magnitude  and the wind will start sweeping and penetrating the CE.

\vspace{-0.6cm}
\bibliographystyle{apj}
\bibliography{../../BibliographyFiles/bibliography}



\end{document}